\documentclass[11pt]{article}

\usepackage{graphicx}
\usepackage{amsmath}
\usepackage{amssymb}
\usepackage{dcolumn}
\usepackage{bm}
\usepackage{authblk}
\usepackage{slashed}
\usepackage{color}
\usepackage[utf8]{inputenc}

\newcommand{\be}{\begin{equation}}
\newcommand{\ee}{\end{equation}}
\newcommand{\ba}{\begin{eqnarray}}
\newcommand{\ea}{\end{eqnarray}}
\newcommand{\no}{\nonumber \\}
\newcommand{\gsim}{\mathrel{\hbox{\rlap{\lower.55ex \hbox {$\sim$}}
                   \kern-.3em \raise.4ex \hbox{$>$}}}}
\newcommand{\lsim}{\mathrel{\hbox{\rlap{\lower.55ex \hbox {$\sim$}}
                   \kern-.3em \raise.4ex \hbox{$<$}}}}

\def\del{\partial}

\def\roughly#1{\mathrel{\raise.3ex\hbox{$#1$\kern-.75em%
\lower1ex\hbox{$\sim$}}}}
\def\lsim{\roughly<}
\def\gsim{\roughly>}

\def\({\left(}
\def\){\right)}
\def\[{\left[}
\def\]{\right]}

\def\l{{\lambda}}

\def\d{{\delta}}

\def\o{{\omega}}
\def\O{{\Omega}}
\def\e{{\epsilon}}

\def\a{{\alpha}}
\def\b{{\beta}}
\def\c{{\chi}}

\def\g{{\gamma}}
\def\G{{\Gamma}}
\def\p{{\pi}}
\def\P{{\Pi}}
\def\m{{\mu}}
\def\n{{\nu}}
\def\r{{\rho}}
\def\s{{\sigma}}

\def\th{{\theta}}
\def\ph{{\phi}}

\def\P{{\Pi}}

\newcommand{\lag}{\langle}
\newcommand{\rag}{\rangle}

\newcommand{\hn}{\hat{n}}

\date{\today}
\begin{document}
\title{\bf Dilepton Helical Production in a Vortical Quark-Gluon Plasma}

\author[1]{Lihua Dong}
\author[1]{Shu Lin}
\affil[1]{School of Physics and Astronomy, Sun Yat-Sen University, Zhuhai 519082, China}

\maketitle

\begin{abstract}
In this paper, we propose an observable counting a weighted difference between right-handed and left-handed lepton pairs, which is coined dilepton helical rate. The weight is the momentum difference of the lepton pairs projected onto an auxiliary vector. We derive the helical rate in a quark-gluon plasma with a vorticity in the limit when the quark and lepton masses are ignored. We find the helical rate is maximized when the auxiliary vector is parallel to the vorticity, in which case it has a nearly spherical oblate ellipsoidal distribution. We propose that it can be used as a vortical-meter for quark-gluon plasma.
\end{abstract}


\newpage

\section{Introduction}

It is believed that off-central heavy ion collisions can produce a strongly vortical fluid \cite{Deng:2016gyh,Jiang:2016woz}. By spin-orbital coupling, the vorticity can lead to spin polarization of final state particles \cite{Liang:2004ph,Liang:2004xn,Becattini:2007sr,Betz:2007kg,Gao:2007bc,Huang:2011ru}. This is indeed confirmed by STAR collaboration in polarization measurement of lambda hyperons \cite{STAR:2017ckg}. While global polarization of lambda hyperon has been nicely understood based on spin-vorticity coupling in thermal models \cite{Becattini:2017gcx,Wei:2018zfb,Wu:2019eyi,Fu:2020oxj,Liu:2019krs}. The local polarization reveals a tension between theoretical predictions and experimental measurements \cite{STAR:2019erd}. There have been significant progress toward understanding the puzzle recently \cite{Fu:2021pok,Liu:2021uhn,Becattini:2021suc,Becattini:2021iol,Yi:2021ryh,Florkowski:2021xvy}. Theoretical frameworks such as quantum kinetic theory \cite{Wang:2019moi,Weickgenannt:2019dks,Gao:2019znl,Li:2019qkf,Hattori:2019ahi,Zhang:2019xya,Liu:2020flb,Guo:2020zpa,Gao:2019zhk,Carignano:2019zsh,Yamamoto:2020zrs,Yang:2020hri,Wang:2020pej,Shi:2020htn,Weickgenannt:2020aaf,Hou:2020mqp,Weickgenannt:2021cuo,Sheng:2021kfc,Wang:2021qnt,Wang:2021owk,Lin:2021mvw,Chen:2021azy} and spin hydrodynamics \cite{Florkowski:2017dyn,Florkowski:2017ruc,Becattini:2018duy,Hattori:2019lfp,Florkowski:2018fap,Bhadury:2020puc,Shi:2020htn,Peng:2021ago,Gallegos:2021bzp,Hongo:2021ona} have been developed for the description of evolution of spin degree of freedom.

The spin polarization of hadrons is tied to the polarized distribution of quarks at freezeout. It is desirable to have a probe sensitive to the polarized quark distribution at early stage of fireball evolution. The electromagnetic probe such as dilepton and photon barely interact with the fireball once produced, thus is a unique probe on the polarized distribution at early stage of the quark-gluon plasma (QGP). There have been extensive studies on polarized dilepton and photon rates arising from hydrodynamic gradient \cite{Dusling:2008xj,Speranza:2018osi}, momentum anisotropy \cite{Ipp:2007ng,Baym:2014qfa,Baym:2017qxy}, magnetic field \cite{Tuchin:2010gx,Tuchin:2012mf,Muller:2013ila,Arciniega:2013dqa,Bandyopadhyay:2016fyd,Wang:2020dsr,Chao:2016ysx,Wang:2021ebh,Wang:2021eud,Das:2021fma}, vorticity \cite{Wei:2021dib} and chiral imbalance \cite{Mamo:2013jda,Mamo:2015xkw}. The last three sources also couple to the spin of quarks, thus can further leads to spin polarized dilepton and photon rates. Indeed, polarization of dilepton and photon can arise from different sources such as magnetic field \cite{Yee:2013qma} and chiral imbalance \cite{Mamo:2013jda,Mamo:2015xkw}.

In this paper, we propose a spin polarized observable for dilepton production induced by vorticity. It is formally defined as
\begin{align}\label{helical}
  \frac{dN(Q)}{d^4Q}=\sum_{P+P'=Q}\(N_R(P,P')-N_L(P,P')\)(P'-P)\cdot \hn,
\end{align}
which is a weighted difference between numbers of right-handed and left-handed lepton pairs\footnote{By right-handed lepton pairs, we refe
r to right-handed fermion and left-handed anti-fermion. Similar terminology applies to left-handed spinors.}. The phase space of the lepton pairs is constrained by $P+P'=Q$, with $P$ and $P'$ being momenta of lepton and anti-lepton, and $Q=(q_0,q\hat{\bf q})$ being momentum of virtual photon with a bin size set by $d^4Q$. The weight $(P'-P)\cdot \hn$ depends on an auxiliary spacelike unit vector $\hn$. The vector $\hn$ can be chosen freely. Since the new observable essentially counts helicity of leptons, we coin it dilepton helical rate\footnote{Here we use helicity of spinor, i.e. both right-handed electron and left-handed positron carry helicity one half.}. As we shall see, the dilepton helical rate has simple angular distribution for generic $q_0$ and $q$. For ${\hat{\bf n}}$ not perpendicular to $\boldsymbol{\o}$, the angular distribution is approximately spherical with
\begin{align}
  \frac{dN(Q)}{d^4Q}\propto \boldsymbol{\o}\cdot\hat{\bf n}.
\end{align}
Therefore the direction that maximizes the dilepton helical rate gives the direction of the vorticity. It can serve as a vortical-meter for QGP.

In heavy ion experiments, dilepton rate is measured by summing over many events. Since the proposed helical rate depends on the orientation of reaction plane, which fluctuates from event to event, a careful summation is needed for this observable. A meaningful summation requires knowledge of orientation of the reaction plane. This is available by the technique employed in \cite{STAR:2017ckg}. Since the proposed helical rate is invariant under rotation, a meaningful summation can be achieved if we simply define $\hat{n}$ with respect to the reaction plane determined for each event.

The paper is organized as follows: in Section~\ref{self-energy}, we calculate the virtual photon self-energy; in Section~\ref{spinsum} and Section~\ref{weighted}, we perform spin sum leading to helicity difference and weighted phase space integrations respectively; the results and phenomenological discussions are presented in Section~\ref{resultp}; we conclude and provide outlook in Section~\ref{conclusion}. Throughout the paper, we ignore the lepton and quark masses.

\section{Photon self-energy in vortical QGP}\label{self-energy}

We start with the $S$-matrix element of dilepton production $(i\to f,l,{\bar l})$ \cite{Bellac:2011kqa}
\begin{align}
S_{fi}(P,P')=-\frac{ie^{2}}{Q^{2}}\bar{u}(P)\g^{\m}v(P')\int d^{4}xe^{iQ\cdot x}\langle f|j_{\m}(x)|i\rangle.
\end{align}
The differential production rate per unit volume is given by
\begin{align}\label{diff_G}
  d\G=\sum_{fi}\frac{1}{\O}|S_{fi}(P,P')|^2\frac{d^3\bold{p}}{2E_p(2\p)^3}\frac{d^3\mathbf{p'}}{2E_p'(2\p)^3},
\end{align}
where $\O$ is the spacetime volume. For massless dilepton, we have $E_p=p$ and $E_p'=p'$.
Using that the initial state is the equilibrium state in the presence of vorticity, we can express the amplitude square as
\begin{align}\label{S2}
  |S_{fi}|^2&=\frac{e^4}{Q^4}\sum_{fi}(\bar{u}(P)\g_{\m}v(P'))(\bar{u}(P)\g_{\n}v(P'))^{*}\int d^4x\int d^4y e^{-iQ\cdot(x-y)}\lag i\vert j_\m(x)\vert f\rag \lag f\vert j_\n(y)\vert i\rag \no
  &=\frac{e^4}{Q^4}\O (\bar{u}(P)\g_{\m}v(P'))(\bar{u}(P)\g_{\n}v(P'))^{*}\P^{\m\n<}(Q),
\end{align}
where $\P^{\m\n<}(Q)=\int d^4x e^{iQ\cdot x}\lag j^\m(0)j^\n(x)\rag$ is to be evaluated in equilibrium with vorticity. We can simplify the differential rate as
\begin{align}\label{pr}
d\G=\frac{e^4}{Q^{4}}(\bar{u}(P)\g_{\m}v(P'))(\bar{u}(P)\g_{\n}v(P'))^{*}\P^{\m\n<}\frac{d^{3}\bold{p}}{2E(2\p)^{3}}\frac{d^{3}\bold{p'}}{2E'(2\p)^{3}}.
\end{align}
Up to now, the spins of the dilepton are arbitrary. A particular spin sum corresponding to the helicity of the lepton pair will be constructed in Section~\ref{spinsum}.

We first calculate the photon self-energy $\P^{\a\b}$ in the equilibrium state with vorticity. We treat the vorticity as a perturbation, which induces a correction to the equilibrium self-energy without vorticity. The lowest order vortical correction can be calculated using following representation of one-loop diagram
\begin{align}\label{Pi1}
\Pi^{\a\b <(1)}(Q)=\int \frac{d^{4}K}{(2\p)^{4}} tr\[\g^{\a}S^{<(1)}(K+Q)\g^{\b}S^{>(0)}(K)+\g^{\a}S^{<(0)}(K+Q)\g^{\b}S^{>(1)}(K)\],
\end{align}
with $S^{<(0)}$ and $S^{<(1)}$ being quark propagators at the zeroth and first order in vorticity. The two terms correspond to vortical correction entering either propagators.
$S^{</>(0)}$ is simply the propagator in local equilibrium
\begin{align}\label{equ}
  S^{<(0)}(K)&=-(2\p)\slashed{K}\d(K^{2})\e(K\cdot u)\tilde{f}(K\cdot u),\no
  S^{>(0)}(K)&=(2\p)\slashed{K}\d(K^{2})\e(K\cdot u)(1-\tilde{f}(K\cdot u)),
\end{align}
where $\e(K\cdot u)$ is the sign function, $u$ is the fluid velocity and $\tilde{f}$ is Fermi-Dirac distribution function
\begin{align}
\tilde{f}(K\cdot u)=\frac{1}{e^{(K\cdot u-\m )/T}+1}.
\end{align}
The vortical correction to propagator $S^{</>(1)}$ can be obtained from solution to Kadanoff-Baym equation in the collisionless limit \cite{Gao:2018jsi,Fang:2016vpj}
\begin{align}\label{offcrc}
S^{<(1)}(K)=S^{>(1)}(K)=-(2\p) \frac{1}{2}\tilde{\slashed{K}}\g^{5}\d(K^{2})\e(K\cdot u)\tilde{f}'(K\cdot u),
\end{align}
with $\tilde{\slashed{K}}=K^{\mu}\tilde{\O}_{\m\n}\g^{\n}$ and $\tilde{\O}_{\m\n}=\frac{1}{2}\e_{\m\n\r\s}\del^{\r}u^{\s}$.
$\tilde{\O}_{\m\n}$ can be decomposed as
\begin{align}
\tilde{\O}_{\m\n}=\o_{\m}u_{\n}-\o_{\n}u_{\m}+\e_{\m\n\r\s}\varepsilon^{\r}u^{\s},
\end{align}
where $\o_{\m}=\frac{1}{2}\e_{\m\n\r\s}u^{\n}\del^{\r}u^{\s}$ and $\varepsilon_{\mu}=\frac{1}{2}u_{\l}\del^{\l}u_{\m}$ corresponding to vorticity and acceleration respectively. We will consider a fluid with vorticity only, so $\varepsilon_\m=0$.

Plugging (\ref{equ}) and (\ref{offcrc}) into \eqref{Pi1}, we find the two terms in \eqref{Pi1} give equal contribution. This can be shown with a change of variable $K\to-K-Q$, under which $\e(k_0+q_0)\e(k_0)$ and $\d((K+Q)^2)\d(K^2)$ are separately invariant and $\tilde{f}'(k_0+q_0)(1-\tilde{f}(k_0))$ is mapped to $\tilde{f}(k_0+q_0)\tilde{f}'(k_0)$. It follows that the two terms are equal, so we can obtain
\begin{align}\label{pi1f}
\Pi^{\a\b <(1)}(Q)&=\frac{4i}{(2\p)^{2}}\int dk_{0}k^{2}dk dcos\th d\phi \ \e(k_{0}+q_{0})\e(k_{0})\d(K^{2})\d((K+Q)^{2})\e^{\a\n\b\l}\tilde{\O}_{\m\n}\nonumber\\
&\times(K+Q)^{\m}K_{\l}
(1-\tilde{f}(k_{0}))\tilde{f}'(k_{0}+q_{0}),
\end{align}
where we have used spherical coordinates for ${\bf k}$ and taken the photon momentum $\bold{q}$ along the z-axis. Here quarks are massless and
$q_{0}$ needs to be positive for the time-like virtual photons to decay into dilepton. Only $k_{0}<0$ and $k_{0}+q_{0}>0$ are kinematically allowed corresponding to the annihilation of the quark-anti-quark pair. In this condition, $\d(K^{2})\d((K+Q)^{2})$ can be simplified as $\frac{\d(k_{0}+k)\d(k_{0}+q_{0}-|\bold{k}+\bold{q}|)}{4k|\bold{k}+\bold{q}|}$. We can use the first and second delta function to calculate the integration of $dk_{0}$ and $dcos\th$ respectively, and then \eqref{pi1f} becomes
\begin{align}\label{pi2f}
\Pi^{\a\b <(1)}(Q)&=-\frac{i}{2\p}(N_{c}\sum_{u,d,s}e^{2}_{q})\int_{(q_0-q)/2}^{(q_{0}+q)/2} \frac{dk}{q}\e^{\a\n\b\l}\tilde{\O}_{\m\n}(K+Q)^{\m}K_{\l}(1-\tilde{f}(k_{0}))\tilde{f}'(k_{0}+q_{0}),
\end{align}
where $k_{0}=-k$ and we have reinserted electric charge factor and number of colors.
Since $\Pi^{\a\b <(1)}$ is antisymmetric in the indices, we decomposed it into $\Pi^{ij<(1)}$ and $\Pi^{0i <(1)}$. We obtain the following nonvanishing components for the vortical correction to the photon self-energy
\begin{align}\label{Pi1_exp}
\Pi^{ij<(1)}(Q)=-\frac{i}{2\p}(N_{c}\sum_{u,d,s}e^{2}_{q})\e^{ijk}\bigg((\boldsymbol{\o}\cdot\hat{\bold{q}})\hat{q}_{k}\mathcal{C}_{1}+\o_{k}\mathcal{C}_{2}\bigg),\nonumber\\
\Pi^{0i<(1)}(Q)=-\frac{i}{2\p}(N_{c}\sum_{u,d,s}e^{2}_{q})\e^{ijk}\hat{q}_{j}\o_{k}\mathcal{C}_{3},
\end{align}
with
\begin{align}\label{chi}
\mathcal{C}_{1}=\mathcal{\chi}_{2}\frac{q^2-3q_{0}^{2}}{2q^{2}}+\mathcal{\chi}_{1}\frac{q_{0}(3q_0^2-q^2)}{2q^2}+\mathcal{\chi}_{0}\frac{(q^2-q_0^2)(q^2+3q_0^2)}{8q^2},\nonumber\\
\mathcal{C}_{2}=\mathcal{\chi}_{2}\frac{q^2+q_{0}^{2}}{2q^{2}}-\mathcal{\chi}_{1}\frac{q_{0}(q_0^2+q^2)}{2q^2}+\mathcal{\chi}_{0}\frac{(q_0^2-q^2)^2}{8q^2},\nonumber\\
\mathcal{C}_{3}=-\mathcal{\chi}_{2}\frac{q_{0}}{q}+\mathcal{\chi}_{1}\frac{3q_{0}^{2}-q^{2}}{2q}-\mathcal{\chi}_{0}\frac{q_{0}(q_{0}^{2}-q^{2})}{2q}.
\end{align}
and
\begin{align}\label{cag2}
\mathcal{\chi}_{n}=\int_{(q_0-q)/2}^{(q_{0}+q)/2}\frac{dk}{q}k^{n}(1-\tilde{f}(-k))\tilde{f}'(-k+q_{0}).
\end{align}

In the limit $q_0\gg T$ and $q_0\gg q$, we can use Boltzmann approximation: $1-\tilde{f}(-k)=\tilde{f}(k)\to e^{-k/T}$ and $\tilde{f}(-k+q_{0})\to e^{-(q_0-k)/T}$. $\mathcal{\c}_n$ can be evaluated analytically and (\ref{chi}) becomes
\begin{align}\label{chiba}
\mathcal{C}_{1}=-\frac{e^{-q_{0}/T}q^2}{6T},\ \
\mathcal{C}_{2}=\frac{e^{-q_{0}/T}(2q_0^2-q^2)}{6T},\ \
\mathcal{C}_{3}=-\frac{q q_{0}e^{-q_{0}/T}}{6T}.
\end{align}



Let us comment on the limit $q\to0$, which seems to lead to divergence. In fact, the divergence does not occur. On the one hand, $\mathcal{\chi}_{n}$ is convergent with the divergent integrand compensated by vanishing integration domain. On the other hand, there will be mutual cancellation among terms in \eqref{chi}, rendering finite results in the end. We confirm this by working out the $q\to0$ limit of $\mathcal{\chi}_{n}$ in appendix A.

 \section{Spin sum for helicity difference}\label{spinsum}

Now we perform spin sum. For unpolarized rate, we simply sum over all spins, which gives the following polarization tensor \cite{Bellac:2011kqa}
\begin{align}\label{l_sum}
l^{\m\n}=\sum\limits_{spins}(\bar{u}(P)\g^{\m}v(P'))(\bar{u}(P)\g^{\n}v(P'))^{*}=4\bigg(P^{\m}P'^{\n}+P^{\n}P'^{\m}-g^{\m\n}(P \cdot P')\bigg),
\end{align}
As we can see, $l^{\m\n}$ is symmetric in indices $\m$ and $\n$, so it vanishes when contracted with anti-symmetric photon self-energy (\ref{Pi1}).
 The fact that vortical correction to unpolarized dilepton rate is not a surprise. Since the vorticity couples to the spin of particles, right-handed and left-handed lepton pairs with the same momenta have opposite spins. The vortical correction cancels among right-handed and left-handed pairs. A nonvanishing correction is expected in the difference of right-handed and left-handed lepton pairs. Below we separate the spin sum for right-handed and left-handed pairs and show this is indeed the case.

We use $L$ and $R$-indices for the spinor helicities. For right-handed lepton pairs, it refers to right-handed fermion and left-handed anti-fermion. Similar terminology applies to left-handed spinors.
Firstly, we work out the right-handed sector $l^{\m\n}_{R}$ explicitly
\begin{align}\label{mnr}
l^{\m\n}_{R}=(\bar{u}_{R}(P)\g^{\m}v_{R}(P'))(\bar{u}_{R}(P)\g^{\n}v_{R}(P'))^{*},
\end{align}
(\ref{mnr}) is a tensor under the Lorentz transformation. In chiral basis, (\ref{mnr}) can be rewritten as
\begin{align}\label{mnr1}
l^{\m\n}_{R}=({u}_{R}^{\dagger}(P)\s^{\m}v_{R}(P'))({u}_{R}^{\dagger}(P)\s^{\n}v_{R}(P'))^{*},
\end{align}
where $\s^{\m}=(1, \boldsymbol{\s})$ and  $\bar{\s}^{\m}=(1, -\boldsymbol{\s})$. Here, $\boldsymbol{\s}$ are the Pauli matrices. $u$ and $v$ are column vectors. We can take advantage of the following representation
\begin{align}\label{se}
u(P)=\frac{1}{\sqrt{2p_{0}}}
\left(
\begin{array}{cc}
P\cdot\s\ \xi\\
P\cdot\bar{\s}\ \xi
\end{array}
\right),\ \ \
v(P')=\frac{1}{\sqrt{2p'_{0}}}
\left(
\begin{array}{cc}
P'\cdot\s\ \eta\\
-P'\cdot\bar{\s}\ \eta
\end{array}
\right).
\end{align}
Two-component spinor $\xi$ and $\eta$ for right-handed spinors satisfy
\begin{align}\label{esp}
\boldsymbol{\s}\cdot\hat{\bold{p}}\ \xi=\xi,\ \ \
\boldsymbol{\s}\cdot\hat{\bold{p}}'\ \eta=\eta.
\end{align}
This causes the upper component of $u$ and $v$ in (\ref{se}) to be zero. However, the nonvanishing lower component gives $u_{R}=\sqrt{2p_{0}}\ \xi$ and $v_{R}=-\sqrt{2p'_{0}}\ \eta$. Then, (\ref{mnr1}) becomes
\begin{align}\label{rlt}
l^{\m\n}_{R}=4p_{0}p'_{0}(\xi^{\dagger}\s^{\m}\eta)(\xi^{\dagger}\s^{\n}\eta)^{*},
\end{align}
We proceed by calculating temporal and spatial components of $\xi^{\dagger}\s^{\m}\eta$ separately. The former is simple
\begin{align}\label{c0}
\xi^{\dagger}\s^{0}\eta=\xi^{\dagger}\eta.
\end{align}
The latter is a complex three vector, which can be decomposed into the following complete basis as
\begin{align}\label{sv1}
\xi^{\dagger}\s^{i}\eta=a p_{i}+ b p'_{i} + c \e^{ijk}p_{j}p'_{k},
\end{align}
The parameter of $a$, $b$ and $c$ can be fixed by dotting (\ref{sv1}) with $p_{i}$, $p'_{i}$ and $\e^{ijk}p_{j}p'_{k}$, which gives
\begin{align}\label{par}
a=\frac{p'_{0}}{p_{0}p'_{0}+\bold{p}\cdot\bold{p'}}\xi^{\dagger}\eta,\ \ \ b=\frac{p_{0}}{p_{0}p'_{0}+\bold{p}\cdot\bold{p'}}\xi^{\dagger}\eta,\ \ \ c=\frac{-i}{p_{0}p'_{0}+\bold{p}\cdot\bold{p'}}\xi^{\dagger}\eta,\ \ \
\end{align}
(\ref{sv1}) can be expressed with (\ref{par}) as
\begin{align}\label{ci}
\xi^{\dagger}\s^{i}\eta=\frac{p'_{0}p_{i}+p_{0}p'_{i}-i\e^{ijk}p_{j}p'_{k}}{p_{0}p'_{0}+\bold{p}\cdot\bold{p'}}\xi^{\dagger}\eta.
\end{align}
$\xi^{\dagger}\eta$ being the overlap of eigenvector of Hamiltonians $\boldsymbol{\s}\cdot\hat{\bold{p}}$ and $\boldsymbol{\s}\cdot\hat{\bold{p}}'$, is still left in (\ref{c0}) and (\ref{ci}). The overlap is supposed to respect rotational symmetry. We can calculate $\xi^{\dagger}\eta$ by going to frame in which ${\bold{p}'}$ is along z axis. Using the Pauli sigma matrices, we can solve the (\ref{esp}) to get $\xi$ and $\eta$ for right-handed as follows
\begin{align}\label{rxe}
\xi=\frac{1}{\sqrt{2p_{0}(p_{0}-p_{3})}}
\left(
\begin{array}{cc}
p_{1}-ip_{2}\\
p_{0}-p_{3}
\end{array}
\right),\ \ \
\eta=
\left(
\begin{array}{cc}
1\\
0
\end{array}
\right).
\end{align}
So it is apparent that $\xi^{\dagger}\eta=(p_{1}+ip_{2})/\sqrt{2p_{0}(p_{0}-p_{3})}$. Obviously the eigenvectors and thus the overlap are defined only up to a phase factors in quantum mechanics. The ambiguity associated with the phase factor is related with side-jump effect \cite{Hidaka:2016yjf}. Fortunately the ambiguity cancels out in the following product
\begin{align}\label{rcf}
(\xi^{\dagger}\eta)(\xi^{\dagger}\eta)^{*}=\frac{p_{1}^{2}+p_{2}^{2}}{{2p_{0}(p_{0}-p_{3})}}=\frac{p_{0}^{2}-(\bold{p}\cdot\hat{\bold{p}}')^{2}}{{2p_{0}(p_{0}-\bold{p}\cdot\hat{\bold{p}}')}}
=\frac{p_{0}p'_{0}+\bold{p}\cdot\bold{p'}}{2p_{0}p'_{0}}.
\end{align}
In the second equality, we have used the fact that ${\bf p'}$ is along $z$ axis to arrive at a manifestly rotational invariant expression.
Combining \eqref{c0}, \eqref{ci} and \eqref{rcf}, we obtain the explicit expressions for components of $l_R^{\m\n}$.
\begin{align}
l^{00}_{R}=2(p_{0}p'_{0}+\bold{p}\cdot\bold{p'}),\nonumber\\
l^{0i}_{R}=2(p'_{0}p_{i}+p_{0}p'_{i}+i\e^{ikl}p_{k}p'_{l}),\nonumber\\
l^{ij}_{R}=\frac{2(p'_{0}p_{i}+p_{0}p'_{i}-i\e^{ikl}p_{k}p'_{l})(p'_{0}p_{j}+ p_{0}p'_{j}+i\e^{jmn}p_{m}p'_{n})}{p_{0}p'_{0}+\bold{p}\cdot\
\bold {p'}}.
\end{align}

Now, we turn to calculate the left-hand sector $l_{L}^{\m\n}$. $l^{\m\n}_{L}$ is also a Lorentz tensor. We write it in chiral basis as
\begin{align}\label{mnl1}
l^{\m\n}_{L}=({u}_{L}^{\dagger}(P)\bar{\s}^{\m}v_{L}(P'))({u}_{L}^{\dagger}(P)\bar{\s}^{\n}v_{L}(P'))^{*},
\end{align}
The expressions of $u$ and $v$ are not changed, whereas, $\xi$ and $\eta$ for left-handed spinors satisfy the following equation which is different from (\ref{esp})
\begin{align}\label{esp1}
\boldsymbol{\s}\cdot\hat{\bold{p}}\ \xi=-\xi,\ \ \
\boldsymbol{\s}\cdot\hat{\bold{p}}'\ \eta=-\eta.
\end{align}
we can obtain $u_{L}=\sqrt{2p_{0}}\ \xi$ and $v_{L}=\sqrt{2p'_{0}}\ \eta$ from the nonvanishing upper components of (\ref{se}). Therefore, (\ref{mnl1}) is simplified to
\begin{align}
l^{\m\n}_{L}=4p_{0}p'_{0}(\xi^{\dagger}\bar{\s}^{\m}\eta)(\xi^{\dagger}\bar{\s}^{\n}\eta)^{*},
\end{align}
Using the same tricks, we can get the temporal and spatial components of $\xi^{\dagger}\bar{\s}^{\m}\eta$
\begin{align}\label{cii}
\xi^{\dagger}\bar{\s}^{0}\eta=\xi^{\dagger}\eta,
\end{align}
\begin{align}
\xi^{\dagger}(-\s^{i})\eta=\frac{p'_{0}p_{i}+p_{0}p'_{i}+i\e^{ijk}p_{j}p'_{k}}{p_{0}p'_{0}+\bold{p}\cdot\bold{p'}}\xi^{\dagger}\eta,
\end{align}
Based on (\ref{esp1}), we obtain $\xi$ and $\eta$ for left-handed spinors as follows
\begin{align}\label{lxe}
\xi=\frac{1}{\sqrt{2p_{0}(p_{0}+p_{3})}}
\left(
\begin{array}{cc}
p_{1}-ip_{2}\\
-p_{0}-p_{3}
\end{array}
\right),\ \ \
\eta=
\left(
\begin{array}{cc}
0\\
1
\end{array}
\right).
\end{align}
Evidently, $\xi^{\dagger}\eta=-(p_{0}+p_{3})/\sqrt{2p_{0}(p_{0}+p_{3})}$. We can arrive at a rotational invariant form as
\begin{align}\label{lcf}
(\xi^{\dagger}\eta)(\xi^{\dagger}\eta)^{*}=\frac{(p_{0}+p_{3})^{2}}{{2p_{0}(p_{0}+p_{3})}}=\frac{p_{0}+p_{3}}{2p_{0}}
=\frac{p_{0}p'_{0}+\bold{p}\cdot\bold{p'}}{2p_{0}p'_{0}}.
\end{align}
It is identical with (\ref{rcf}). Consequently, the left-handed sectors $l^{\m\n}_{L}$ are obtained by making use of (\ref{cii}) and (\ref{lcf})
\begin{align}
l^{00}_{L}=2(p_{0}p'_{0}+\bold{p}\cdot\bold{p'}),\nonumber\\
l^{0i}_{L}=2(p'_{0}p_{i}+p_{0}p'_{i}-i\e^{ikl}p_{k}p'_{l}),\nonumber\\
l^{ij}_{L}=\frac{2(p'_{0}p_{i}+p_{0}p'_{i}+i\e^{ikl}p_{k}p'_{l})(p'_{0}p_{j}+ p_{0}p'_{j}-i\e^{jmn}p_{m}p'_{n})}{p_{0}p'_{0}+\bold{p}\cdot\
\bold {p'}}.
\end{align}
It is easy to verify $l_R^{\m\n}+l_L^{\m\n}$ reproduces $l^{\m\n}$ in \eqref{l_sum}. We are interested in their difference
\begin{align}\label{asn}
h^{\m\n}=l^{\m\n}_{R}-l^{\m\n}_{L},
\end{align}
which is expected to have nonvanishing response to vorticity.
It is easy to get
\begin{align}\label{llt}
h^{00}=0,\nonumber\\
h^{0i}=4i\e^{ijk}p_{j}p'_{k},\nonumber\\
h^{ij}=4i\e^{ijk}(p'_{0}p_{k}-p_{0}p'_{k}),
\end{align}
In arriving at $h^{ij}$, we have used Schouten identity $\e^{ijk}V_{l}-\e^{jkl}V_{i}+\e^{kli}V_{j}-\e^{lij}V_{k}=0$ in three dimensional. (\ref{llt}) implies that $h^{\m\n}$ is anti-symmetric in indices $\m$ and $\n$. In fact, \eqref{llt} adopts a manifestly Lorentz invariant form
\begin{align}\label{ldiff}
h^{\m\n}=4i\e^{\m\n\r\s}P_{\r}P'_{\s}.
\end{align}

\section{Weighted phase space integration}\label{weighted}
In order to obtain the dilepton rate \eqref{pr}, we should integrate $h_{\m\n}$ with phase space. Denoting the integrated result by $H_{\m\n}$, we have
\begin{align}\label{hmn}
H_{\m\n}=\int\frac{d^{3}\bold{p}}{2E(2\p)^{3}}\frac{d^{3}\bold{p'}}{2E'(2\p)^{3}}(2\p)^{4}\d^{4}(Q-P-P')h_{\m\n}.
\end{align}
Unfortunately, $H_{\m\n}$ vanishes identically because $h_{\m\n}$ is odd under the exchange $P\leftrightarrow P'$ while the remainder of the integrand and integration measure are even under the exchange. In order to arrive at a nonvanishing result, we need to make the integrand odd under the exchange. This is possible if we introduce an additional weight $(P'-P)\cdot \hat{n}$ in the phase space integration:
\begin{align}\label{Hmn}
H_{\m\n}=\int\frac{d^{3}\bold{p}}{2E(2\p)^{3}}\frac{d^{3}\bold{p'}}{2E'(2\p)^{3}}(2\p)^{4}\d^{4}(Q-P-P')\bigg((P'-P)\cdot \hat{n}\bigg) h_{\m\n},
\end{align}
where $\hat{n}$ is a unit auxiliary vector, which can be chosen freely. This choice has the desired property under exchange and at the same time is easy to implement in experiment. It is simply the momentum difference between lepton $u({\bf p})$ and anti-lepton $v({\bf p}')$ projected onto the $\hat{n}$ axis.

Since $H_{\m\n}$ is anti-symmetric in indices, we only need to calculate $H_{0i}$ and  $H_{ij}$. The former can be calculated as
\begin{align}\label{H0i}
H_{0i}=\frac{i\e_{ijk}}{2\p}\int\frac{p^{2}dcos{\th}dp}{p|\bold{q}-\bold{p}|}\d(q_{0}-p-|\bold{q}-\bold{p}|)\bigg((2P-Q)\cdot \hat{n}\bigg)p_{j}(q_{k}-p_{k}),
\end{align}
where we have used spherical coordinates for ${\bf p}$ and taken the photon momentum $\bold{q}$ along the z-axis. We have also used $E= p$ and $E'= p'$ for massless lepton pairs. We can use the delta function in (\ref{H0i}) to evaluate the integration of $dcos\th$ to give
\begin{eqnarray}
H_{0i}&=&\frac{i\e_{ijk}}{2\p}\int_{(q_0-q)/2}^{(q_{0}+q)/2}\frac{dp}{q}\bigg((2P-Q)\cdot \hat{n}\bigg)(p_{j}q_{k}-p_{j}p_{k})\nonumber\\
&=&\frac{iQ^2}{12\pi}\e_{ijk}q_{j}\hat{n}_{k},
\end{eqnarray}
$H_{ij}$ can be determined by using the same method.
\begin{eqnarray}\label{Hij}
H_{ij}&=&\frac{i\e_{ijk}}{2\p}\int_{(q_0-q)/2}^{(q_{0}+q)/2}\frac{dp}{q}\bigg((2P-Q)\cdot \hat{n}\bigg)(p_{0}p'_{k}-p'_{0}p_{k})\nonumber\\
&=&\frac{iQ^2}{12\pi}\e_{ijk}(q_{0}\hat{n}_{k}-q_{k}\hat{n}_{0}).
\end{eqnarray}
(\ref{H0i}) and (\ref{Hij}) can be combined into a compact expression
\begin{align}\label{H_compact}
H_{\m\n}=-\frac{iQ^2}{12\pi}\e_{\m\n\r\s}Q^{\r}\hat{n}^{\s}.
\end{align}

\section{Results and Phenomenological discussion}\label{resultp}

Finally, we can give the ``helical differential rate'' for lepton pair in terms of the vortical correction to self-energy $\Pi^{\a\b <(1)}$ \eqref{Pi1_exp} and weighted tensor $H^{\m\n}$ \eqref{H_compact}
\begin{eqnarray}\label{result}
\frac{dN}{d^{4}Q}=\frac{e^{4}}{Q^{4}}\frac{1}{(2\pi)^4}(N_{c}\sum_{u,d,s}e^{2}_{q})H_{\m\n}\Pi^{\m\n <(1)},
\end{eqnarray}
\eqref{result} simply counts the difference between number of right-handed and left-handed lepton pairs, weighted by the projected momentum difference between the leptons in each pair. Its discrete form is given by \eqref{helical} in the introduction.
We stress that the weight has mass dimension one, so the weighted differential rate $dN$ has one more mass dimension than the conventional differential rate $d\G$.
For convenience, we decompose \eqref{result} into the following form
\begin{eqnarray}\label{Hp0i}
H_{0i}\Pi^{0i<(1)}&=&\frac{Q^{2}q}{24\p^{2}}\mathcal{C}_{3}(\boldsymbol{\o}\cdot\hat{\bold{n}}-(\hat{\bold{q}}\cdot\hat{\bold{n}})(\hat{\bold{q}}\cdot\boldsymbol{\o})),\nonumber\\
H_{ij}\Pi^{ij<(1)}&=&\frac{Q^{2}}{12\p^{2}}((\boldsymbol{\o}\cdot\hat{\bold{n}})q_{0}\mathcal{C}_{2}+(\hat{\bold{q}}\cdot\hat{\bold{n}})(\hat{\bold{q}}\cdot\boldsymbol{\o})q_{0}\mathcal{C}_{1}).
\end{eqnarray}
Here, we have chosen the temporal component of spacelike vector $\hat{n}$ to be zero, namely, $\hat{n}_{0}=0$. Then plugging \eqref{Hp0i} into \eqref{result}, we have
\begin{eqnarray}\label{Result}
\frac{dN}{d^{4}Q}&=&\frac{e^{4}}{Q^{4}}\frac{1}{(2\pi)^4}(N_{c}\sum_{u,d,s}e^{2}_{q})(2H_{0i}\Pi^{0i<(1)}+H_{ij}\Pi^{ij<(1)})\nonumber\\
&=&\frac{\a^{2}}{Q^{2}}\frac{1}{12\pi^{4}}(N_{c}\sum_{u,d,s}e^{2}_{q})\bigg((\boldsymbol{\o}\cdot\hat{\bold{n}})R_{1}+(\hat{\bold{q}}\cdot\hat{\bold{n}})(\hat{\bold{q}}\cdot\boldsymbol\o)R_{2}\bigg).
\end{eqnarray}
with
\begin{align}\label{r1ar2}
R_{1}=q_{0}\mathcal{C}_{2}+q\mathcal{C}_{3},\nonumber\\
R_{2}=q_{0}\mathcal{C}_{1}-q\mathcal{C}_{3}.\nonumber\\
\end{align}
In the Boltzmann approximation, (\ref{r1ar2}) can be calculated analytically by using (\ref{chiba}) as follows:
\begin{align}\label{br1ar2}
R_{1}=\frac{e^{-q_{0}/T}}{3T}q_{0}(-q^{2}+q_{0}^{2}),\ \ \ R_{2}=0.
\end{align}
\begin{figure}[htbp]
\centering
\includegraphics[width=0.44\textwidth]{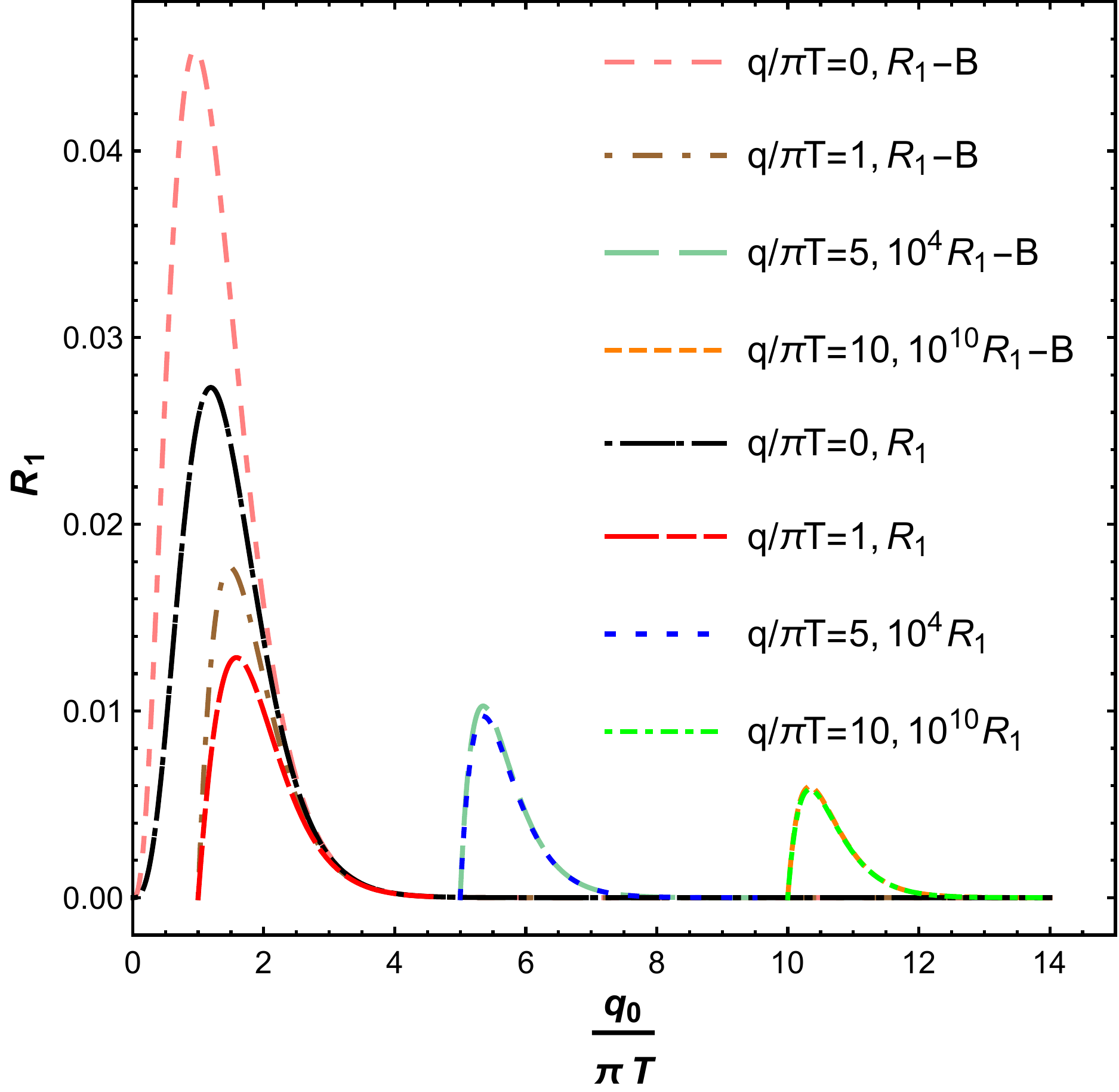}
\includegraphics[width=0.46\textwidth]{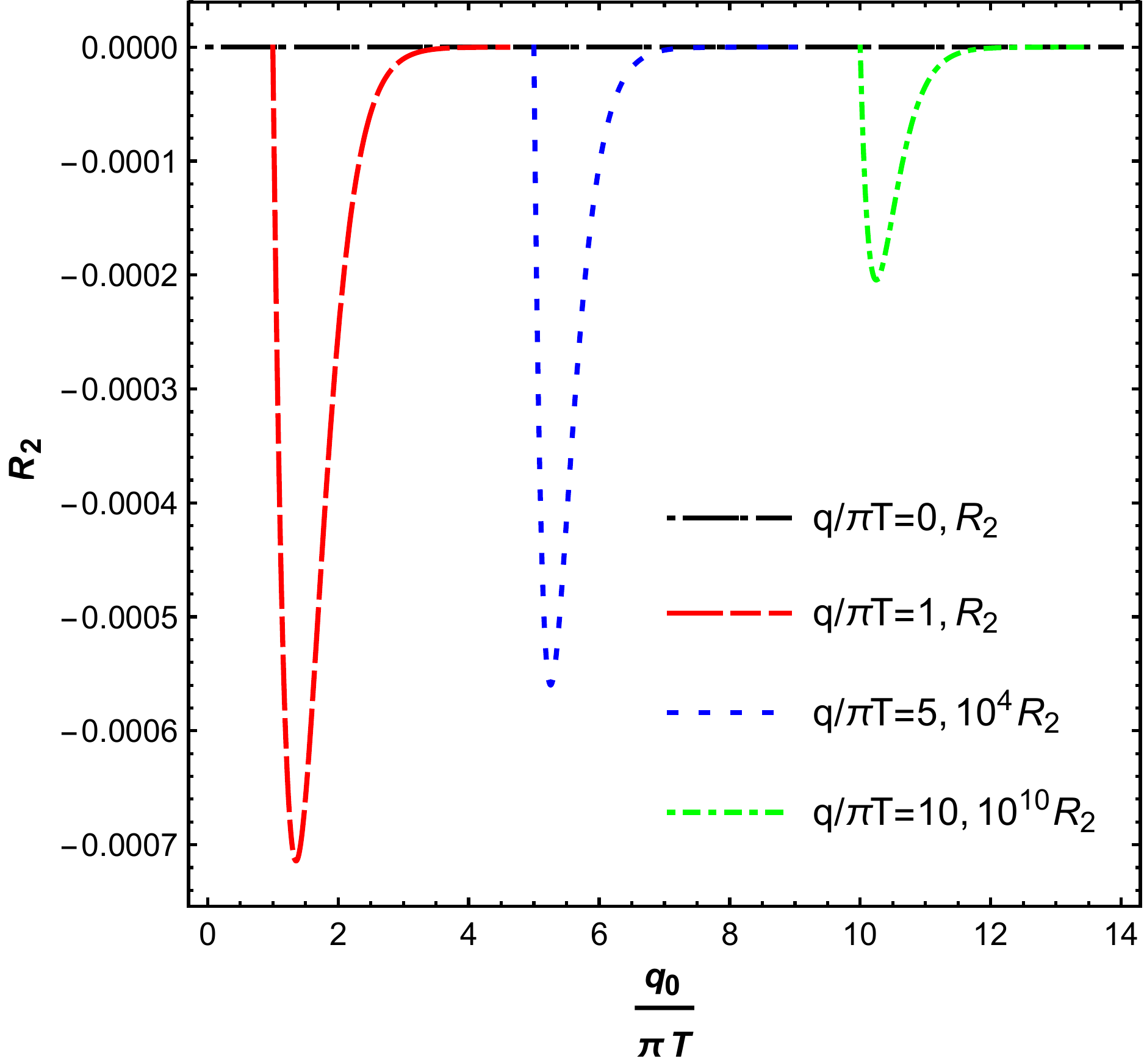}
\vspace*{0.1cm}
\caption{Left: The $q_{0}$ dependence of positive $R_{1}$ with different $q/\p T$.  $R_{1}-B$ represents the result of $R_1$ in the Boltzmann approximation.  Right: The $q_{0}$ dependence of negative $R_{2}$ with different $q/\p T$. At large $q_0$, the Boltzmann approximation becomes accurate and at small $q_0$, the Boltzmann approximation tends to overestimate the exact results. For generic choice of $q_0$ and $q$, $R_{1}$ is about twenty times that of $R_{2}$}\label{R1}
\end{figure}
According to (\ref{r1ar2}) and (\ref{chi}), we can see that $R_{1}$ and $R_{2}$ are rotational invariant functions. We show them as a function of $q_0$ for different $q$ in Fig.~\ref{R1} together with the Boltzmann approximated results \eqref{br1ar2}.  Since $R_{2}$ vanishes in the Boltzmann approximation, we do not show it any more in Fig.~\ref{R1}. We see at large $q_0$, the Boltzmann approximation becomes accurate and at small $q_0$, the Boltzmann approximation tends to overestimate the exact results. Moreover, we find $R_{1}$ and $R_{2}$ are positive and negative respectively for generic choice of $q_0$ and $q$, and $R_{1}$ is about twenty times that of $R_{2}$.

The angular distribution of the helical differential rate is encoded in the factors $(\boldsymbol{\o}\cdot\hat{\bold{n}})$ and $(\hat{\bold{q}}\cdot\hat{\bold{n}})(\hat{\bold{q}}\cdot\boldsymbol\o)$.
It is instructive to consider three special cases of $\hat{\bold{n}}$: i. $\hat{\bold{n}}$ parallel to $\boldsymbol{\o}$, the helical rate is dominated by $R_{1}$ giving a spherical angular distribution. The correction from the $R_{2}$ deforms the sphere into an approximate oblate ellipsoid. The rate is positive with more right-handed lepton pairs than left-handed ; ii. $\hat{\bold{n}}$ anti-parallel to $\boldsymbol{\o}$, both terms change sign and helical rate is still an oblate ellipsoid. iii. $\hat{\bold{n}}$ perpendicular to $\boldsymbol{\o}$, we can take $\hat{\bold{n}}$ and $\boldsymbol{\o}$ along $x$ and $y$ axis respectively, the resulting angular distribution $\propto \sin^2\th\cos\ph\sin\ph$.
\begin{figure}[htbp]
\centering
\includegraphics[width=0.45\textwidth]{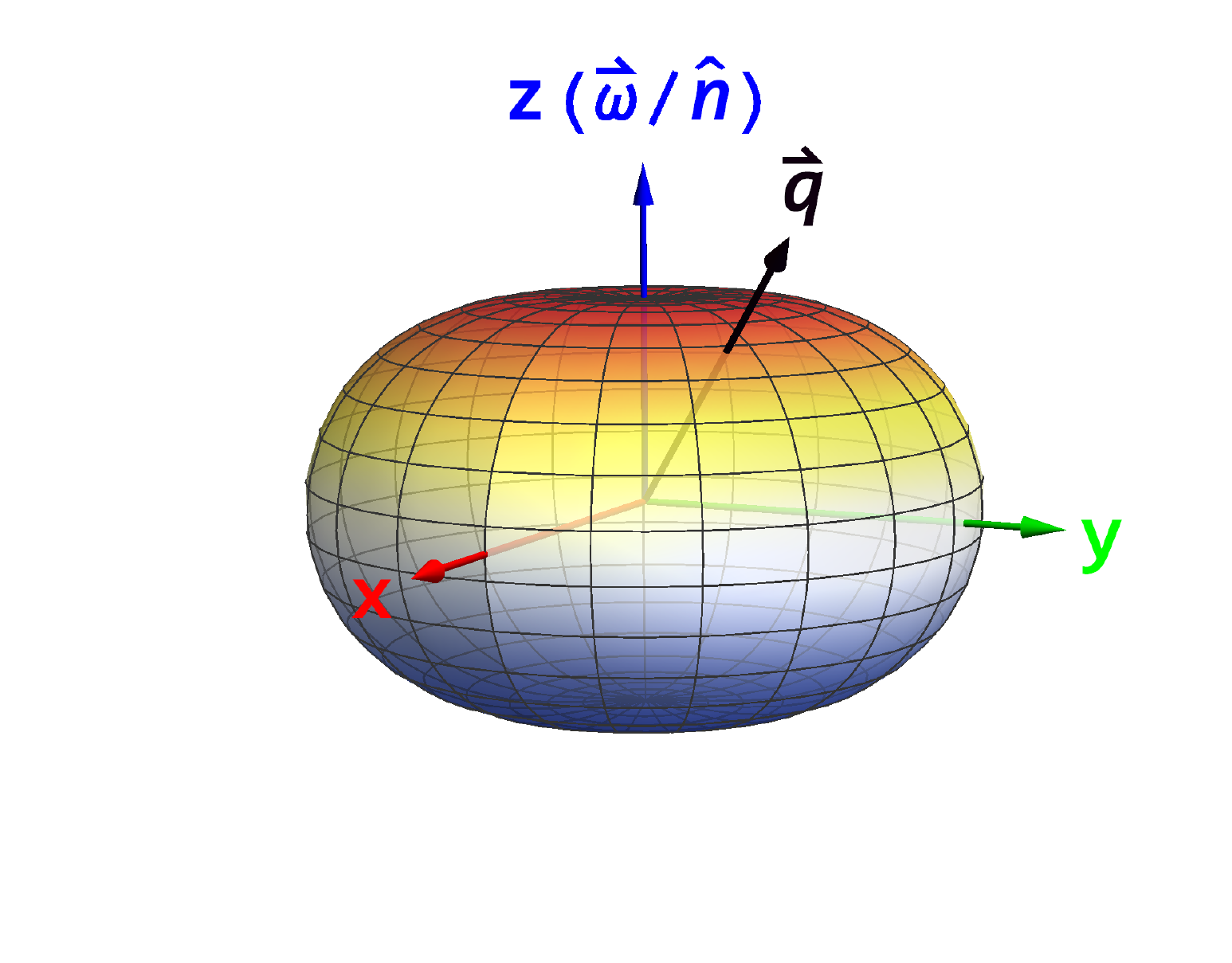}
\includegraphics[width=0.48\textwidth]{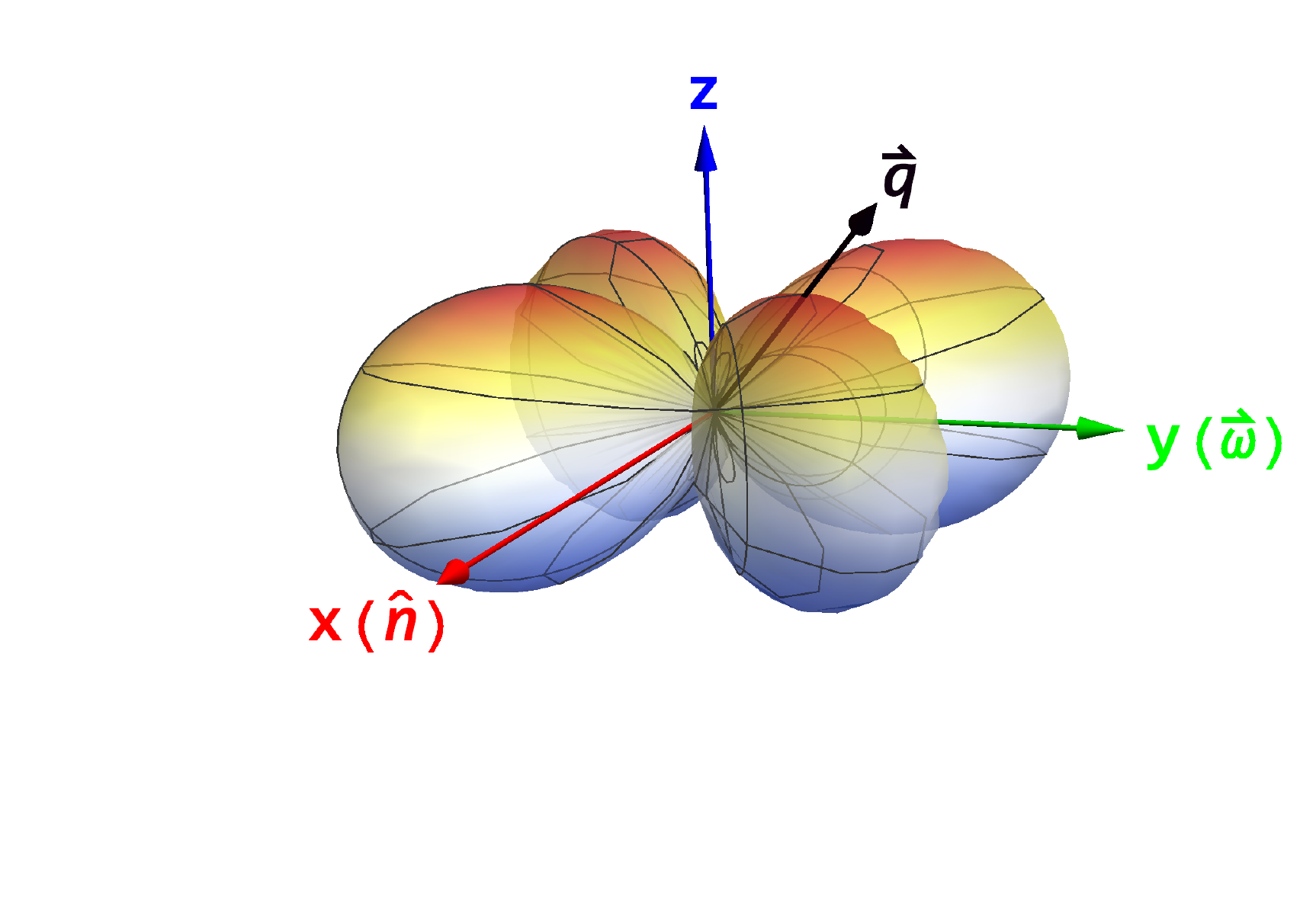}
\vspace*{0.1cm}
\caption{The schematic plots of angular distribution of the dilepton helical rate. Left: $\hat{\bold{n}}$ parallel to $\boldsymbol{\o}$ along $z$ axial, $R_{1}$ gives a spherical angular distribution and the correction from $R_{2}$ deforms the sphere into an approximate oblate ellipsoid. Right: $\hat{\bold{n}}$ perpendicular to $\boldsymbol{\o}$, $\hat{\bold{n}}$ and $\boldsymbol{\o}$ are along $x$ and $y$ axis respectively. The resulting angular distribution $\propto \sin^2\th\cos\ph\sin\ph$.}\label{aRth}
\end{figure}

The fact that the helical rate maximizes when $\hat{\bold{n}}$ is parallel to the vorticity is most easily understood in the extreme case $q=0$. In this case $(P'-P)\cdot\hn=2\bold{p}\cdot\hat{\bold{n}}$ with $\bold{p}$ being momentum of lepton. For right-handed (left-handed) lepton, it is proportional to (minus) the spin projected onto $\hat{\bold{n}}$. Since spin tends to align with vorticity, it follows that the helical rate from the difference between right-handed and left-handed lepton pairs maximizes when $\hat{\bold{n}}$ is parallel to vorticity.
Two schematic plots of the angular distribution of the dilepton helical rate for the case of i and iii are shown in Fig.~\ref{aRth}. The fact the leading term is spherically symmetric is very useful. As long as ${\bf\hat{n}}$ is not exactly perpendicular to $\boldsymbol{\o}$, the helical rate is always maximized when the auxiliary vector is parallel to the vorticity, so that it can be used as a vortical-meter for the QGP.

Let us comment on a phenomenological aspect of \eqref{Result}. The dependence of \eqref{Result} on the vorticity indicates that the observable is sensitive to the orientation of reaction plane. In reality, the low rate of thermal dilepton production requires summing data over many events, for which the orientation of reaction plane fluctuates. In order to perform a meaningful summation, we need to have knowledge of the reaction plane. The orientation of reaction plane in each event can be estimated by the technique in \cite{STAR:2017ckg}. Since the observable we propose is invariant under rotation, a meaningful summation can be achieved if we define $\hat{n}$ with respect to the reaction plane in each event.

\begin{figure}[htbp]
\centering
\includegraphics[width=0.3\textwidth]{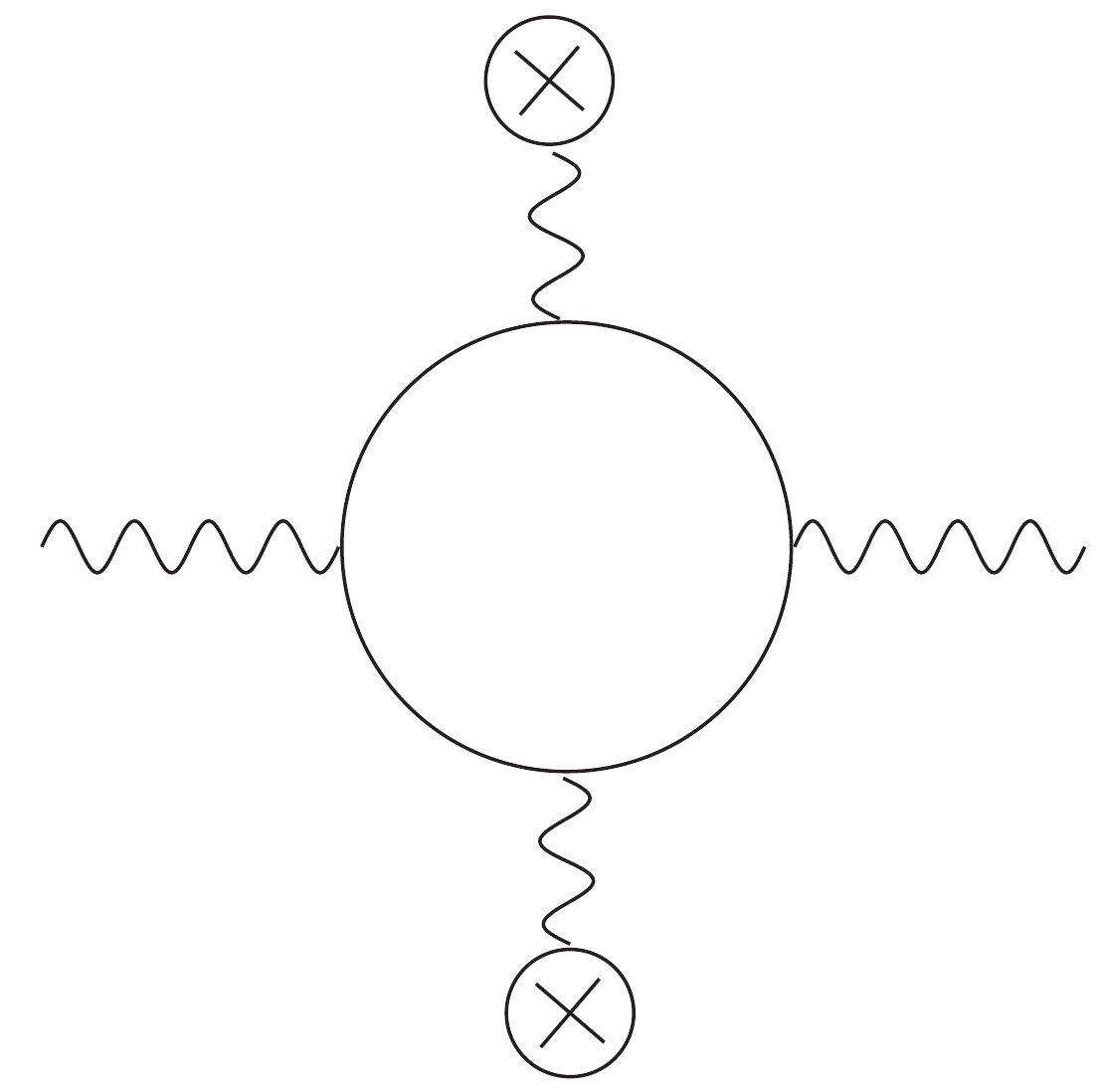}
\vspace*{0.1cm}
\caption{One-loop photon self-energy diagrams with two background magnetic field insertions. Two other diagrams with two background magnetic field inserted in a same propagator are not shown. None of them lead to anti-symmetric components of photon self-energy.}\label{fm1}
\end{figure}

Before closing this section, we address the distinguishability of the vortical contribution to the proposed observable. Indeed one would expect sources such as magnetic field and chiral imbalance to contribute to the same observable as well. Since the chiral imbalance fluctuates from event to event independent of the orientation of the reaction plane, the contribution from chiral imbalance simply sums to zero.
For the magnetic field contribution, we now show it also vanishes at one-loop order. The reason is the photon self-energy in a background magnetic field contains no anti-symmetric component, which is needed for the helical rate. The photon self-energy in a background magnetic field arises from summation of one-loop self-energy diagrams with arbitrary numbers of background field insertions, as illustrated in Fig.~\ref{fm1}. In neutral plasma we consider, diagrams with odd number of magnetic external lines vanish by Furry's theorem, and diagrams with even number of magnetic external lines only lead to symmetric components by the fermion trace. It is worth pointing out though anti-symmetric components of self-energy do arise in higher loop diagrams, see for example \cite{Wang:2021ebh}, which effectively incorporate dissipative effect. Those contributions are suppressed by additional powers of coupling constant.

\section{Conclusion and Outlook}\label{conclusion}

We have proposed an observable counting a weighted difference between right-handed and left-handed lepton pairs, which is coined dilepton helical rate. We found the helical rate is sensitive to the vorticity of QGP. It is maximized when the auxiliary vector is parallel to the vorticity, thus it can serve as a vortical-meter for QGP. In the special cases when the auxiliary vector is parallel or anti-parallel to the vorticity, the angular distribution of the helical rate has a shape of a nearly spherical oblate ellipsoid.

It would be desirable to extend the analysis to the hadronic phase, where the measurement of spin polarization of vector meson has been undertaken \cite{Singha:2020qns}. It would be interesting to find out if the spin polarization of vector meson can have an imprint on the helical rate proposed in this study.
It would also be interesting to include the collision effect in the present study. In the presence of collision, the polarization of quarks are known to have an additional component orthogonal to vorticity due to the polarization rotation effect \cite{Hou:2020mqp}. It would be interesting to find out the phenomenological consequence of it.

\section*{Acknowledgments}
We are grateful to Xu-Guang Huang for useful discussions. This work is in part supported by NSFC under Grant Nos 12075328, 11735007 and 11675274.

\appendix

\section{Limits of $\mathcal{C}_n$ and $\c_n$ at small $q$}
We are interested in the behavior of $\mathcal{C}_{n}$ in the limit where $q$ tends to zero.
We start with the following small $q$ expansions of $\c_n$, which are easily obtained from \eqref{cag2}
\begin{align}\label{mathchi}
\chi_{2}=-\frac{q_{0}^{2}e^{q_{0}/2T}}{4T(1+e^{q_{0}/2T})^{3}}-\frac{ q^{2}e^{q_{0}/2T}(8+8e^{q_{0}/T}-\frac{8q_{0}}{T}+\frac{q_{0}^{2}}{T^{2}}+e^{q_{0}/2T}(16-\frac{8q_{0}}{T}-\frac{3q_{0}^{2}}{T^{2}}))}
{96T(1+e^{q_{0}/2T})^{5}}+\mathcal{O}[q]^{4},\nonumber\\
\chi_{1}=-\frac{q_{0}e^{q_{0}/2T}}{2T(1+e^{q_{0}/2T})^{3}}+\frac{ q^{2}e^{q_{0}/2T}(4-\frac{q_{0}}{T}+e^{q_{0}/2T}(4+\frac{3q_{0}}{T}))}
{48T^{2}(1+e^{q_{0}/2T})^{5}}+\mathcal{O}[q]^{4},\nonumber\\
\chi_{0}=-\frac{e^{q_{0}/2T}}{T(1+e^{q_{0}/2T})^{3}}+\frac{ q^{2}e^{q_{0}/2T}(-1+3e^{q_{0}/2T}))}
{24T^{3}(1+e^{q_{0}/2T})^{5}}+\mathcal{O}[q]^{4}.
\end{align}
these results in (\ref{mathchi}) clearly display that all the terms in the expansion are finite and even in $q$. The finiteness of the expansion is a consequence of cancellation between divergent integrand and vanishing integration domain.
We only need to consider the contribution given by the leading order and the next leading order terms of $\chi_{n}$. Plugging \eqref{mathchi} into \eqref{chi}, we find the following limits for $\mathcal{C}_n$
%
\begin{align}\label{mathc}
\mathcal{C}_{1}(q\to{0})=0,\ \ \ \
\mathcal{C}_{2}(q\to{0})=\frac{1}{3}\frac{q_{0}^{2}e^{q_{0}/2T}}{T(1+e^{q_{0}/2T})^3},\ \ \ \
\mathcal{C}_{3}(q\to{0})=0.
\end{align}
Plugging (\ref{mathc}) into (\ref{r1ar2}), we can obtain
\begin{align}\label{br1ar2s}
R_{1}=\frac{q_{0}^{3}e^{q_{0}/2T}}{3T(1+e^{q_{0}/2T})^{3}},\ \ \ R_{2}=0.
\end{align}
which have been used in Fig.~\ref{R1} at $q/\p T =0$.

\end{document}